# Stripping syntax from complexity: An information-theoretical perspective on complex systems

Rick Quax[1,*], Omri Har-Shemesh[1], Stefan Thurner[4,5,6], Peter M.A. Sloot[1,2,3]

1. Computational Science, University of Amsterdam, Science Park 904, 1098 XH Amsterdam, The Netherlands
2. Complexity Institute, Nanyang Technological University, 60 Nanyang View, Singapore 639673, Republic of Singapore
3. National Research University of Information Technologies, Mechanics and Optics (ITMO), Kronverkskiy 49, 197101 Saint Petersburg, Russia
4. Section for Science of Complex Systems; Medical University of Vienna, Spitalgasse 23; A-1090, Austria
5. Santa Fe Institute; 1399 Hyde Park Road; Santa Fe; NM 87501; USA.
6. IIASA, Schlossplatz 1, A-2361 Laxenburg; Austria.

* Corresponding author: r.quax@uva.nl

**Abstract.** Claude Shannon's information theory (1949) has had a revolutionary impact on communication science. A crucial property of his framework is that it decouples the meaning of a message from the mechanistic details from the actual communication process itself, which opened the way to solve long-standing communication problems. Here we argue that a similar impact could be expected by applying information theory in the context of complexity science to answer long-standing, cross-domain questions about the nature of complex systems. This happens by decoupling the domain-specific model details (e.g., neuronal networks, ecosystems, flocks of birds) from the cross-domain phenomena that characterize complex systems (e.g., criticality, robustness, tipping points). This goes beyond using information theory as a non-linear correlation measure, namely it allows describing a complex system entirely in terms of the storage, transfer, and modification of informational bits. After all, a phenomenon that does not depend on model details should best be studied in a framework that strips away all such details. We highlight the first successes of information-theoretic descriptions in the recent complexity literature, and emphasize that this type of research is still in its infancy. Finally we sketch how such an information-theoretic description may even lead to a new type of universality among complex systems, with a potentially tremendous impact. The goal of this perspective article is to motivate a paradigm shift in the young field of complexity science using a lesson learnt in communication science.

**Popular Summary.** Seemingly very different complex systems may actually display the same kind of behavior. For instance, 'criticality' is a useful type of instability to rapidly respond to external stimuli in the brain, gene-regulatory networks, bacterial colonies, and the flocking of birds, among others. From this the picture emerges that the mechanistic details of complex systems may be irrelevant for generating such cross-domain phenomena as criticality. However, the young field of complexity science currently lacks a unified framework to study emergent phenomena without recourse to studying mechanistic details of particular complex systems, which complicates the matter and limits the rates of scientific progress. Information theory has already solved a similar problem in communication science a few decades ago, for which it was originally created. Here we argue that information theory – or an appropriately generalized version – may bring a similar revolution in the form of a unified theoretical framework in complexity science.

# 1. Introduction

During most of the existence of the telephone the natural paradigm of transmitting a voice through a telephone line was to directly convert sound waves into electric waves. Long-distance communication became possible by repeatedly amplifying these waves. Noiseless, error-free communication would be achieved if the electric waves remained unperturbed along the line during transmission. However, in practice noise substantially interfered with the electric waves and telephone engineers struggled to understand and solve this issue.

In hindsight one could argue that the engineering problem of reliable, error-free telephone communication was hindered by inadvertently entangling two notions: syntax and semantics. The syntax of a message is the encoding of the message (electric waves), whereas the semantics refers to the meaning of it. This entanglement is manifested by directly converting sound waves into electric waves and by subsequently solving the technical transportation problem through noisy channels which would perturb the signals' waveforms. The difficulty in solving an entangled problem of this kind is evidenced, in retrospect, by the suboptimal handling of crippling noise in communication for many decades (1, 2) preceding the digital era.

# 2. Separating syntax from semantics

Then in 1948, Claude Shannon formulated a framework named 'information theory' (3) which effectively separated the syntax from the semantics of a communication event. Shannon argued that for the engineering problem of transmitting a message it is irrelevant whether the message is a spoken message, a picture, or a sequence of symbols—much less the meaning of this message to the human receiver, referred to as "psychological factors" by Hartley (4). The only relevant concept is how much information there is to be transmitted.

In Shannon's view, a communication channel from $A$ to $B$ is abstracted to a stochastic process in which the state in $A$ induces a state in $B$, where $A$ and $B$ are stochastic variables. As a simple example, think of an early telegraph where the opening and closing of an electrical circuit at $A$ changes a needle's position at $B$. This process can be represented as a conditional probability distribution $\Pr(B|A)$. Once the sender has chosen his message $A=a$ from his a priori distribution $\Pr(A)$, the communication mechanism induces a state in $B$ with the probability $\Pr(B=b|A=a)$. The task for the receiver is to observe $B=b$ and then infer the intended message $A=a$. The receiver's uncertainty can be represented by the inverted conditional probability $\Pr(A|B)$. In case of noiseless communication, $\Pr(A=a|B=b)=1$, whereas for maximally noisy communication, $\Pr(A|B)=1/N$, with $N$ being the total number of possible messages.

# 3. The information viewpoint

Shannon's essential innovation is the notion that the sender possesses a certain amount of *information* (namely the fact that $A=a$) and that this information will be transmitted, such that eventually the receiver will obtain the same information that the sender had. It does not matter what this information is (semantics); what matters is that the sender's information becomes the same as that of the receiver.

For example, if the sender has only two possible states ('yes' or 'no') then the amount of information stored in the sender's state equals 1 yes/no question. Learning the answer to a single yes/no question is sufficient to identify the state of the sender. In units of information we say that the sender stores one *bit*. In case that the sender has four possible states, then its information equals 2 bits. In general, the average amount of information stored in the sender's state $A$ is determined by its marginal probability distribution $\Pr(A)$. It is known as its Shannon entropy,

$$H(A) = -\sum_a \Pr(A=a) \log \Pr(A=a).$$

After a perfect, noiseless transmission, the receiver's information will share exactly $H(A)$ bits with the information stored at the sender. After a failed transmission the receiver shares zero information with the sender, and for noisy transmission their *mutual information* is somewhere in between. It is quantified by

$$I(A:B) = \sum_{a,b} \Pr(A=a, B=b) \log \frac{\Pr(A=a, B=b)}{\Pr(A=a)\Pr(B=b)}.$$

Returning to the example of the telephone, the sender's voice is a waveform which represents a certain amount of information (number of bits) needed to identify the waveform. The receiver must receive this amount of information to be able to infer the sender's waveform, but this does not imply that the waveform itself must actually be transmitted. The idea of decoupling the *encoding* of a message from the message itself may be arguably the largest conceptual innovation in information theory, and one which is perhaps also the most difficult to fully appreciate today since it is now such a widespread notion. Nevertheless its impact on communication theory has been substantial. One of the tremendous consequences is that it is possible to show that there exists a particular encoding strategy for a given message, such that it can be transmitted error-free even though the channel is noisy, as long as the information production rate of the source is less than the channel capacity. The channel capacity is the maximum number of bits per second that a given 'line' can transmit. This is a property of the channel *itself*—independent of the particular encoding chosen for a signal. In fact, the table is now turned for an optimal communication flow: an encoding must be designed in order to approach the channel capacity as closely as possible. Other domains where this decoupling led to fundamental insights include cryptography and data compression.

## 4. Networks of information transmissions

The above example can be extended straightforwardly to communication *networks*. Suppose that at a given moment in time $N$ sender-receiver pairs $A_i, B_i$ communicate over the same (telephone) network. The network must maximize the mutual information $I(A_i : B_i)$ between sender-receiver pairs, while minimizing the 'interference' $I(A_i : B_j), i \neq j$, between other pairs (assuming independence among the messages). This also extends to bi-directional communication. Other than these informational constraints, the choice of the particular implementation of the network is irrelevant: any system that satisfies these constraints can be considered a suitable and efficient communication network.

## 5. Complex networked systems

We argue that a similar decoupling of syntax and semantics in complex systems may well lead to a boost in complexity science, similar in spirit to that of communication theory. In other words, by describing a system solely in terms of how it stores, transmits, and modifies information may lead to new methods to answer cross-domain questions. We argue that this route might be possible even though systems considered by Shannon and complex systems may be radically different in terms of ergodicity and extensibility.

Different scientific domains aim to solve different, domain-specific models to understand systemic behaviors or make predictions. For example, a network of neurons is usually studied by solving (or simulating) a neurobiological model of electric potentials, thresholds and conductance, such as the FitzHugh–Nagumo model (5). At a different spatiotemporal level, each biological cell internally regulates its activity through its gene-regulatory network, usually modelled as protein concentrations (expression levels) which up- or down-regulate the expression of other genes (6) leading to coupled ODEs or reaction-diffusion systems. In the yet different domain of statistical physics, the magnetization of a material is often modelled as a (usually regular) network of spin-spin couplings, where each spin attempts to minimize its local energy induced by the orientation of other spins (7). Other superficially disparate examples include the flocking behavior of birds, social opinion-formation, pattern-formation in traffic flows, and synchronization in financial derivatives markets.

Despite the differences in terminology and the detailed mechanisms in various systems across the scientific domains, the types of questions that researchers attempt to understand are often strikingly similar. Examples include: what is the resilience of a system? Which nodes are most influential? How robust is the system to removal, or to external fluctuations? Is the system metastable, or does it exhibit criticality? Will it pass through a tipping point? What is the role of network topology?

Naturally it would be tremendous progress if such questions could be universally answered without particular domain knowledge. It may prove helpful to phrase and answer these questions within one single theoretical framework. In other words, the domain-specific mechanistic details of a model may be irrelevant for answering cross-domain questions that typically address systemic properties of systems. If such 'universality' truly exists then solving a model which does include mechanistic details and specific domain knowledge makes the task of addressing systemic questions awkward and unnecessarily difficult. The knowledge of details will not lead to additional quality or benefit for the answer of the problem.

## 6. From different domain-specific models to a single information-theoretic language

Here we see a parallel with communication science and Shannon's information theory. Suppose that we are able to formulate each cross-domain question in the language of information theory – similar to the mutual information constraint between sender and receiver in communication. Suppose further that we can translate each mechanistic system description into a single information-theoretic description – similar to the situation where the channel capacity characterizes any communication transmission mechanism. Then, it stands to reason that this approach could lead to significant advances in the understanding of complex networked systems, similar to the unambiguous, well-documented advances in communication theory (1, 2).

First signs that such an approach might be successful are already emerging in the complexity science literature. For instance, regarding the question mentioned above 'which nodes are most influential?', the authors R.Q. and P.M.A.S. recently defined the most influential node to be the one whose information 'lives longest' in the network, i.e., continues to be onward transmitted (8). This led to the seemingly paradoxical prediction that, for a wide class of networked dynamical systems, it is not the hubs that are the most influential nodes for determining the system's short-term behavior. In financial derivatives data they measured a similar measure of 'information lifetime' and demonstrated that it peaks at the time of Lehman Brother's collapse (9), suggesting that it might play an important role in systemic resilience and maybe tipping points.

In fact, it appears that the field of complexity science is currently witnessing the dawn of a widespread growth of using information theory to describe and understand the behavior of complex (networked) systems. To name a few illustrative examples, Lizier et al. propose a framework to formulate dynamical systems in terms of distributed 'local' computation: information storage, transfer, and modification (10) defined by individual terms of Eq. (2). For cellular automata they demonstrate that so-called particle collisions are the primary mechanism for local modifying information, and for a networked variant they show that a phase transition is characterized by the shifting balance of local information storage over transfer (11). Williams and Beer trace how task-relevant information flows through a minimally cognitive agent's neurons and environment to ultimately be combined into a categorization decision (12) or sensorimotor behavior (13). How local interactions lead to multi-scale systemic behavior is also a domain which benefits from information-theoretic approaches, such as by Bar-Yam et al. (14, 15) and Kristian Lindgren (16). Finally, in order to extend information theory itself to handle complexity, multiple authors are concerned with decomposing a single information quantity into multiple constituents, such as synergistic information, including Crutchfield *et al.* (17, 18), Williams and Beer (19), Olbrich et al. (20), and Griffith and Koch (18).

## 7. Universality from information

At the more fundamental level we envision that the information-theoretic description might help to uncover a new 'universality' between seemingly different systems. Universality is the phenomenon known in physics where completely different statistical systems behave 'the same way' under certain circumstances, irrespective of the details of their underlying microscopic dynamics. The best known universality is that of systems near a continuous phase transition where the exact scaling behavior of many quantities depends only on the number of dimensions and the symmetries of the system, and not on the exact form of the Hamiltonian. Many different systems with the same number of dimensions and symmetries will share exactly the same scaling exponents that determine the systemic behavior usually at phase transitions (21). However, the concept of universality is much broader. As an example, a recently introduced form of universality was designed for a certain class of networked systems, grouping them in terms of their response to small perturbations (22).

In an information theoretic view, universality can be thought of being induced by a 'many-to-one' mapping of mechanistic system descriptions to their information-theoretic counterparts. This means that many different system descriptions will lead to one and the same information-theoretic description. As a consequence many cross-domain questions (see also Fig. 1) that can be expressed entirely in terms of information must have the same answer in all these systems, regardless of their mechanistic details.

To see how universality is induced, consider the mutual information between two stochastic variables $A$ and $B$ as part of the information-theoretic description. The probability distributions

$\Pr(A)$ and $\Pr(B|A)$ are high-dimensional objects; for instance, in the simplest case that each variable is a single bit, their joint probability distribution is parameterized by a point in a three-dimensional unit cube. The mutual information function maps this high-dimensional construct to a single real number, introducing a potentially large reduction of dimensionality.

The same is true for the entropy functional. In fact, the entropy functional reveals an additional source of many-to-one mapping, apart from dimension reduction, namely from symmetry inside the dimensions. For instance, if $A$ is one bit then $\Pr(A)$ is parameterized by a single real number on the unit line. The same is true for its entropy $H(A)$ so dimensionality is not reduced, however both $\Pr(A) = p$ and $\Pr(A) = 1 - p$ lead to the same entropy for every $p$, effectively introducing a symmetry around $p = 1/2$.

It remains an open question what a complete information-theoretical description should look like, and whether such a description will be unique or if multiple variants will exist. Building blocks to identify universality that is induced as depicted above, might include networks, combinatorics, and symmetries. It will also be important to handle non-ergodicity and non-extensivity which complex systems may possess. First steps to generalize entropy in this direction have been taken (23, 24). Nevertheless, finding a universality for cross-domain concepts would be a tremendous advance in the science of complex systems. Speculating further, such universality may even encode the very notion of 'stripping the syntax' to find only the relevant features that answer a cross-domain question.

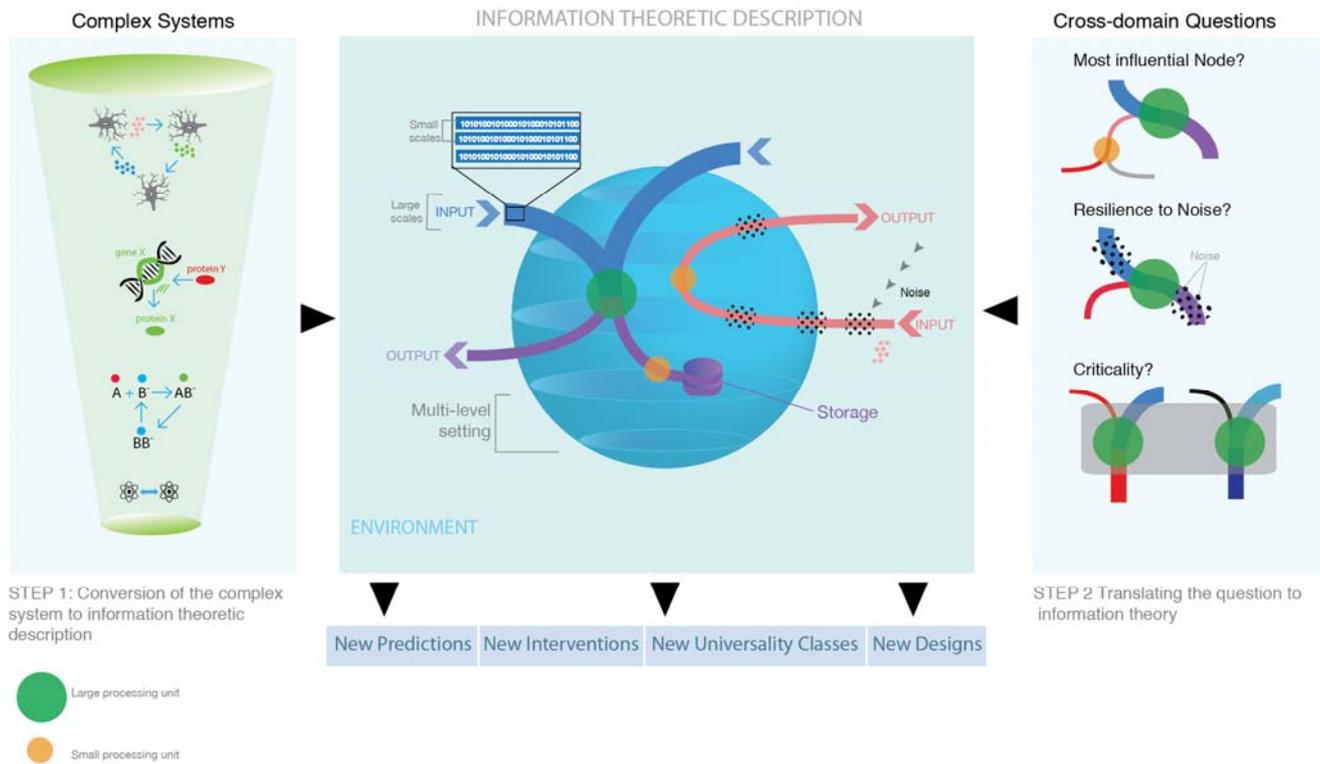

**Figure 1.** Schematic diagram of the envisioned research agenda of using appropriately generalized information theory to address cross-domain questions in complex systems science. In a first step a system's mechanistic description is translated into an informational description, such as how informational bits are stored, transferred, and modified in the system (left panel). Next, research questions must be translated into appropriate information-theoretic counterparts, such as the example of the 'lifetime' of information to measure a node's impact mentioned in the main text (right panel). A potential outcome is a new type of universality among complex systems (Section 7). Another potential outcome could be to find novel universal answers to cross-domain questions that enhance and complement the corresponding state-of-the-art, domain-specific research (Section 8).

# 8. Example: an information-theoretic description to explain criticality

A particular successful example of using a single information-theoretic description to better understand a cross-domain phenomenon starts with the observation that the brain self-organizes towards 'criticality'. This is a statistical physics notion for the fact that it operates at the boundary between different types of dynamics (25), and often is associated with typical scale-free statistics of e.g. activity bursts.

It was subsequently conjectured that the purpose of the brain's criticality is to optimize the information processing capabilities of the brain. Namely, on the one hand the brain must *store* information by persisting in an attractor, but at the same time it must make rapid context switches by transitioning between attractors, leading to information *processing*. The brain must therefore position itself on the 'edge' between these two types of operations in order to rapidly switch between them.

Some researchers then turned this reasoning around in recent work (26, 27). For instance, Hidalgo *et al.* start from information-theoretic constraints and demonstrate that any adaptive system which copes with changing environments will evolve towards such a critical point. Indeed, this criticality phenomenon has been observed in various biological systems with similar information processing constraints (28) and at different scales, ranging from gene-regulatory networks (encoding the decision-making of a single cell) and bacterial colonies (29), to flocking of birds (collectively responding rapidly to external stimuli). In fact a picture emerges where indeed the mechanistic details of the systems seem irrelevant for the emergence of (in this case) criticality, and that information theory links it to adaptability, responsiveness, and stability.

## 9. A note on generalizing Shannon's information theory

In the above we focused on Shannon's classical information theory as the basis for the elusive framework of characterizing how complex systems process information. However, for certain classes of systems this may be insufficient. For instance, some strongly interacting systems –in fact most complex systems- are known to violate assumptions for which Shannon's information theory is built, in particular the 4$^{th}$ Khinchin axiom (30) that ensures ergodicity in classical information theory. In such cases a generalization of the information theory is needed to cope with an essential feature of complexity. One possibility of such generalizations are so-called *generalized entropies* (23, 24). Nevertheless, the essential points above remain invariant under these generalizations, such as the universality induced by dimensionality reduction and symmetries.

## 10. Concluding remarks

Our goal for this discussion is to suggest a coherent perspective and motivation on the use of appropriately generalized information theory to study important questions in complex networked systems. Inspired by the revolutionary impact of information theory on communication science we believe that a similar revolution is possible for complexity science.

We emphasize that this approach is still in its infancy. Many open and fundamental questions remain to be answered. How should any given mechanistic description be translated (encoded) into an information-theoretic description? How can a mechanistic research question be translated into its information-theoretic counterpart? For which types of questions does this approach work, and for which does it fail? Is there one single information-theoretic framework suited for all systems and questions, or will there be multiple frameworks for different questions? Will it be easier to 'solve' an information-theoretic system description as opposed to a mechanistic (PDE, agent-based, etc.) description? How does the information-theoretic framework induce universality classes on the mechanistic descriptions, and how can they be exploited, how can they be characterized – what defines the universality classes?

A wide variety of novel research is currently performed within and around these questions, both theoretical and applied. The information-theoretic techniques vary from Shannon and Fisher information to algorithmic information; the concrete applications vary from a pattern-forming reaction-diffusion model to real financial systems; and the theoretical endeavors include modifications and adaptations of the axioms of Shannon's information theory as well as inferring (the limits of) causality in probabilistic models. Each individual contribution is a valuable advance in its own right. However, an overarching comprehensive framework is not yet obvious from their ensemble, indeed reflecting the current state of the field itself.

**Acknowledgments**

PMAS and RQ acknowledge the financial support of the Future and Emerging Technologies (FET) program within Seventh Framework Programme (FP7) for Research of the European Commission, under the FET- Proactive grant agreement TOPDRIM, number FP7-ICT-318121. PMAS, RQ, and OHS also acknowledge the financial support of the Future and Emerging Technologies (FET) program within Seventh Framework Programme (FP7) for Research of the European Commission, under the FET-Proactive grant agreement Sophocles, number FP7-ICT-317534. PMAS acknowledges the support of the Russian Scientific Foundation, Project #14-21-00137. ST acknowledges the support of the FP7 projects LASAGNE and MULTIPLEX.

**Conflicts of Interest**

The authors declare no conflict of interest.